\documentclass[prl,reprint,showpacs,superscriptaddress,longbibliography]{revtex4-1} 
\usepackage{hyperref} 
\hypersetup
{
    colorlinks=true,       
    linkcolor=blue,        
    citecolor=blue,        
    urlcolor=blue           
}
\usepackage{amsmath}
\usepackage{amssymb} 
\usepackage{bm}
\usepackage{graphicx} 
\usepackage{subfigure} 
\usepackage{dcolumn} 
\usepackage{color}
\usepackage{textcomp} %
\usepackage{ulem}

\begin{document}
\title{Chiral anomaly in noncentrosymmetric systems induced by spin-orbit coupling}  
\author{Suik Cheon}
\affiliation{Department of Physics, Pohang University of Science and Technology, Pohang 37673, Korea}  
\author{Gil Young Cho} 
\affiliation{Department of Physics, Pohang University of Science and Technology, Pohang 37673, Korea}
\affiliation{Center for Artificial Low Dimensional Electronic Systems, Institute for Basic Science (IBS), Pohang 37673, Korea}  
\affiliation{Asia Pacific Center for Theoretical Physics, Pohang 37673, Korea}  
\author{Ki-Seok Kim}
\affiliation{Department of Physics, Pohang University of Science and Technology, Pohang 37673, Korea}
\affiliation{Asia Pacific Center for Theoretical Physics, Pohang 37673, Korea}   
\author{Hyun-Woo Lee}
\email{hwl@postech.ac.kr}
\affiliation{Department of Physics, Pohang University of Science and Technology, Pohang 37673, Korea}
\date{\today}
\begin{abstract}  
The chiral anomaly may be realized in condensed matter systems with pairs of Weyl points. Here we show that the chiral anomaly can be realized in diverse noncentrosymmetric systems even without Weyl point pairs when spin-orbit coupling induces nonzero Berry curvature flux through Fermi surfaces. This motivates the condensed matter chiral anomaly to be interpreted as a Fermi surface property rather than a Weyl point property. The spin-orbit-coupling-induced anomaly reproduces the well-known charge transport properties of the chiral anomaly such as the negative longitudinal magnetoresistance and the planar Hall effect in Weyl semimetals. Since it is of spin-orbit coupling origin, it also affects the spin transport and gives rise to anomaly-induced longitudinal spin currents and the magnetic spin Hall effect, which are absent in conventional Weyl semimetals.
\end{abstract}
%
\maketitle 
%
%
%

{\it Introduction.---}
Massless relativistic fermions with pairs of Weyl points have the chiral symmetry, which may be broken by quantum fluctuations in the presence of electric and magnetic fields ${\bf E}$ and ${\bf B}$, resulting in the chiral anomaly~\cite{PhysRev.177.2426,Nielsen:1983ce},
%
\begin{equation}\label{Eq:chiral_anomaly}
    \frac{\partial n_{\chi} }{\partial t} + \nabla \cdot {\bf J}_{\chi} 
    =
    \chi \frac{e^2}{4 \pi^2  \hbar^2 c} ( {\bf E} \cdot {\bf B}), 
\end{equation}
where $n_{\chi}$ and ${\bf J}_{\chi}$ are respectively the number density and the current density of fermions with the the chiral charge or the chirality $\chi(=\pm 1)$.  
$e ~ (<0)$ is the electron charge.  
The chiral anomaly was introduced originally in high energy physics to explain the anomalous decay of neutral pions.
Recently, it received much attention in condensed matter physics since electron spectra in Weyl semimetals~\cite{RevModPhys.90.015001} resemble those of massless relativistic fermions with pairs of Weyl points and possess the chiral symmetry. 
The chiral anomaly may be realized in such condensed matter systems~\cite{bevan1997momentum} and result in interesting transport phenomena such as the negative longitudinal magnetoresistance~\cite{PhysRevLett.109.181602,PhysRevB.88.104412,Burkov:2015JPCM,PhysRevB.91.245157} and the planar Hall effect~\cite{PhysRevB.96.041110,PhysRevLett.119.176804}, which are verified in experiments~\cite{PhysRevLett.111.246603,Xiong:2015kl,PhysRevX.5.031023,li2016chiral,PhysRevB.97.201110,PhysRevB.98.041103}.

On the other hand, there is controversy regarding the relation between these transport phenomena and the chiral anomaly since these phenomena may occur in systems that are {\it not} Weyl semimetals~\cite{PhysRevB.95.165135,PhysRevLett.119.166601,PhysRevLett.120.026601,PhysRevB.98.081202,Nandy:2018fb,PhysRevB.101.125203,pal2021berry}.
Some of such systems have the helical symmetry and the associated helical charge obeys the same anomalous conservation equation~\cite{PhysRevLett.120.026601,PhysRevB.104.L241111} as Eq.~\eqref{Eq:chiral_anomaly}.
The resulting helical charge pumping can give rise to the aforementioned transport properties.
In this letter, we generalize the chiral anomaly concept further that applies to any systems with nonzero Berry curvature flux through Fermi surfaces, regardless of the energy dispersion shape (relativistic or nonrelativistic), the number of relevant Weyl points (four, two, zero~\cite{PhysRevLett.120.026601,PhysRevB.104.L241111}, or even one), or symmetries of Hamiltonian (helical symmetry or not).
As a specific example, we demonstrate the generalized chiral anomaly for noncentrosymmetric spin-orbit-coupled (SOC) systems with nonrelativistic fermions and only one relevant Weyl point. 
In addition to the aforementioned charge transport phenomena, they exhibit interesting anomaly-induced {\it spin transport} phenomena since the anomaly arises from the SOC.

\begin{figure}[t!]     
  \includegraphics[width=8cm,height=5.744cm]{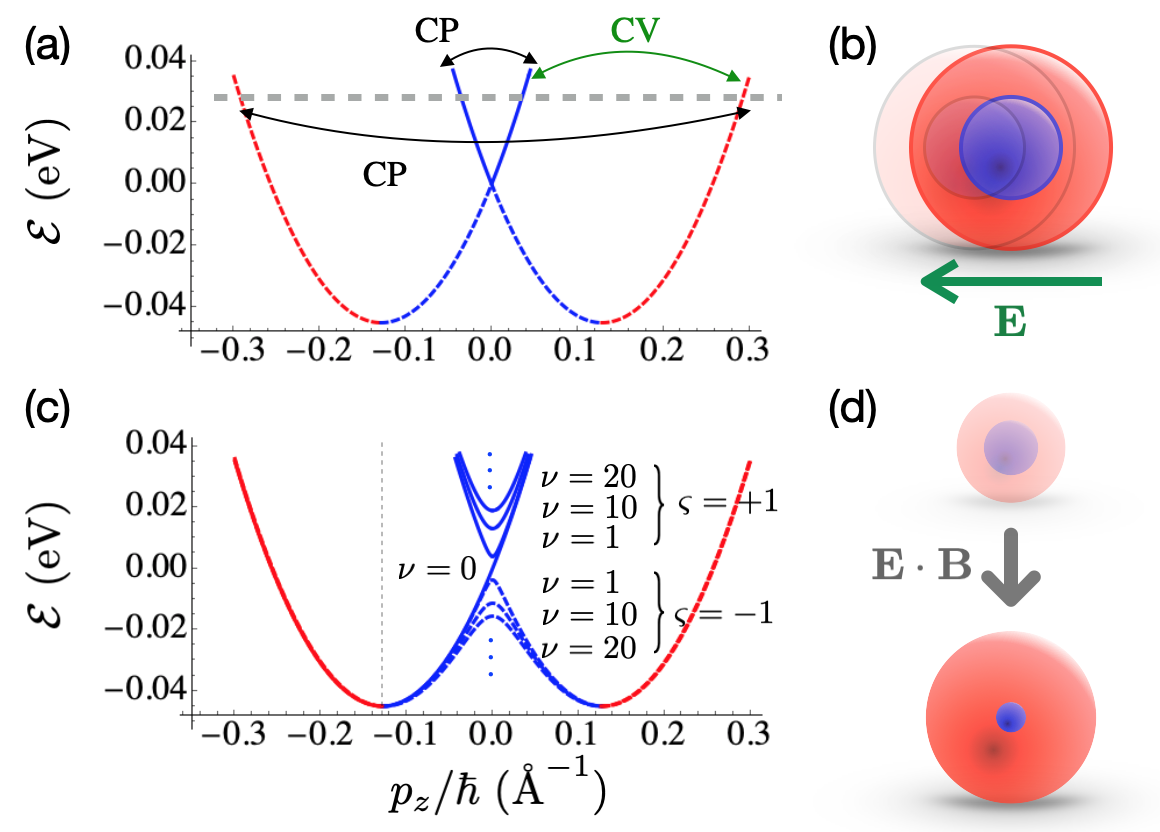}
   \subfigure{\label{fig:Metal_B0}}
   \subfigure{\label{fig:FSduetoE}}
   \subfigure{\label{fig:Metal_LLbandweakB}}
   \subfigure{\label{fig:EdotBFS}} 
\caption 
   {%
   (a) The energy dispersion for a Weyl SOC system without the ${\bf B}$ field.
   The red (blue) line represents states connected to the outer (inner) Fermi surface with the chirality $\chi = +1$ $(-1)$.
   On the other hand, the solid (dashed) line represents the upper (lower) band with the helicity band index $\varsigma = +1$ $(-1)$.
   The grey horizontal dashed line denotes the Fermi energy $\mathcal{E}_{\rm{F}}$.
   Black and green arrows represent the chirality-preserving (CP) scattering and the chirality-violating (CV) scattering processes, respectively. 
   (b) The red (blue) sphere represents the Fermi surface $\chi = +1$ $ (-1)$.
   When an electric field ${\bf E}$ is applied, the ground state Fermi surface (transparent ones) are shifted.
   (c) Landau level dispersion for a Weyl SOC system with $B = 1$ T along the $z$-axis.
   (d) Schematic illustration of the $\chi = +1$ $( \chi = -1)$ Fermi surface expansion (shrinkage) due to the chiral anomaly-induced pumping.
   Parameters are $m^* = 1.4 m_e$, and $\lambda = 0.7$ eV$\cdot$\AA, which are relevant to K$_2$Sn$_2$O$_3$~\cite{He:2019wn}.
   }
\label{fig01}
\end{figure}

{\it SOC-induced chiral anomaly.---}
For illustration, we take a simple Hamiltonian
\begin{equation}\label{Eq:iso_Hamiltonian}
     \mathcal{H}   = \frac{  {\bf p}^2 }{2 m^*} + \frac{\lambda}{\hbar} {\bf p} \cdot \bm{\sigma}, 
\end{equation}
where $m^*$ is an effective electron mass, $\lambda$ is the SOC parameter, and $\bm{\sigma}$ is the Pauli matrix vector for electron spin. 
We take both $m^*$ and $\lambda$ to be positive for concreteness but these constraints may be lifted. 
Figure~\ref{fig:Metal_B0} shows the energy dispersion of $\mathcal{H}$. 
Although the two-fold spin degeneracy is lifted by the SOC, which is a relativistic effect, $\mathcal{H}$ basically has massive nonrelativistic dispersion and there is only one Weyl point at the time-reversal-invariant point ${\bf k}={\bf 0}$.
Thus $\mathcal{H}$ does not have the chiral symmetry.
It can be verified that $\mathcal{H}$ does not have the helical symmetry either~\cite{PhysRevLett.120.026601,PhysRevB.104.L241111}. 
One remark is in order. Due to the Fermion doubling theorem~\cite{NIELSEN198120,NIELSEN1981173}, there should exist another Weyl point somewhere, either far away in the $\mathbf{k}$-space or far away in the energy range where $\mathcal{H}$ does not hold. 
Such systems with distant Weyl points are termed Kramers-Weyl fermion~\cite{Chang:2018bb}.
But properties (such as location and dispersion) of the other Weyl point do not affect the results presented below and in this sense, $\mathcal{H}$ describes a situation where only Weyl point is relevant. 
$\mathcal{H}$ with the Weyl-type SOC structure ${\bf p} \cdot \bm{\sigma}$ can be realized in the noncentrosymmetric point groups {\bf T} and {\bf O}~\cite{KVSamokhin:2009dg} such as K$_2$Sn$_2$O$_3$, $\beta$-RhSi, CoSi, and AlPt~\cite{Chang:2018bb,He:2019wn,PhysRevLett.122.076402,Rao:2019ha,Sanchez:2019es,ter:2019kf}. 

We check the chiral anomaly of $\mathcal{H}$ in three ways: via the Berry curvature flux, semiclassical analysis, and quantum analysis.
For the first analysis, we recall Ref.~\cite{PhysRevLett.109.181602}, which reports that the nonvanishing Berry curvature flux through a Fermi surface is an indicator of the chiral anomaly.
Although Ref.~\cite{PhysRevLett.109.181602} appears to assume Weyl-semimetal-type systems implicitly~\cite{PhysRevB.88.104412}, its result is intriguing since the Berry curvature flux can be calculated regardless of the presence of the chiral symmetry.
Interestingly, $\mathcal{H}$ has two Fermi surfaces [Fig.~\ref{fig:FSduetoE}] and the Berry curvature flux $k_{\iota}$ through the outer ($\iota= +1$)/ inner ($\iota = -1$) Fermi surface,
\begin{equation}\label{Eq:Chern1}
	k_{\iota} = \frac{1}{2 \pi \hbar} \oint_{\text{F.S.} \iota} d {\bf S}_{\bf p} \cdot {\bf \Omega}_{\varsigma\bf p}, 	
\end{equation}
is $+1$/$-1$ for the Fermi energy $\mathcal{E}_{\rm{F}} >0$ and $+1$/$+1$ for $\mathcal{E}_{\rm{F}} < 0$.
For conventional Weyl semimetals with a pair of Weyl points, the sum of $k_{\iota}$'s for the two Fermi surfaces is zero, which contrasts with the result for $\mathcal{H}$ with $\mathcal{E}_{\rm{F}}<0$.
In Eq.~\eqref{Eq:Chern1}, the integration runs over the momentum space area on the Fermi surface $\iota$ (F.S.$\iota$) with the momentum space area element $d {\bf S}_{\bf p}$ pointing outwared.
${\bf \Omega}_{\varsigma {\bf p}} = i \hbar \langle \nabla_{\bf p} u_{\varsigma {\bf p} } | \times |  \nabla_{\bf p} u_{\varsigma {\bf p} } \rangle$ is the Berry curvature~\cite{RevModPhys.82.1959} at the momentum ${\bf p}$ for the upper ($\varsigma = +1$) or lower ($\varsigma = -1$) energy band that lies on the Fermi surface $\iota$. 
To understand the implication of the nonzero Berry curvature flux through the Fermi surfaces, we attempt to derive the ``Fermi surface" version of the chiral anomaly by assigning the chiral index $\chi$ for each eigenstate of $\mathcal{H}$ as follows: $\chi = +1$ $(-1)$ for each eigenstate of $\mathcal{H}$ if the state is ``connected" to the outer (inner) Fermi surface.
To clarify what is meant by ``connected", states with $\chi = +1$ $(-1)$ are marked red (blue) in Fig.~\ref{fig:Metal_B0}.
Note that we now use $\chi$ as a {\it Fermi surface} index rather than the Weyl point index.
This allows $n_{\chi}$ and ${\bf J}_{\chi}$ to be defined unambiguously for $\chi = +1$ and $-1$ even though $\mathcal{H}$ has only one Weyl point.
Using the semiclassical equations of motion in the presence of ${\bf E}$ and ${\bf B}$,
\begin{eqnarray}
	\dot{\bf r}
	&=& {\bf v}_{\varsigma{\bf p}} + \dot{\bf p} \times {\bf \Omega}_{\varsigma{\bf p}}, 
	\label{Eq:eomr}
	\\
	\dot{\bf p}
	&=& e {\bf E} + \frac{e}{c} \dot{\bf r} \times {\bf B},	
	\label{Eq:eomp}
\end{eqnarray}
one can verify after tedious calculation~\cite{supp} that $n_{\chi}$ and ${\bf J}_{\chi}$ defined this way satisfy Eq.~\eqref{Eq:chiral_anomaly} except that the index $\chi$ on the right-hand side of Eq.~\eqref{Eq:chiral_anomaly} is replaced by $\mathcal{C}_{\chi}$,
\begin{equation}
	\mathcal{C}_{\chi} =  \frac{1}{2 \pi \hbar} \oint_{\text{F.S.} \chi} d {\bf S}_{\varsigma {\bf p} } \cdot {\bf \Omega}_{\varsigma\bf p}, 	
\end{equation}
which is $+1$/$-1$ for $\chi =$$+1$/$-1$ {\it regardless} of $\mathcal{E}_{\rm{F}}$.
$\mathcal{C}_{\chi}$ differs from $k_{\chi}$ [Eq.~\eqref{Eq:Chern1}] in that the momentum space area element $d {\bf S}_{\varsigma {\bf p} }$ is pointing along the group velocity direction, which is {\it inward} for the inner Fermi surface with $\mathcal{E}_{\rm{F}}<0$, and outward otherwise.
Thus this semiclassical analysis supports the chiral anomaly for $\mathcal{H}$, provided that the chirality index $\chi$ is redefined as the Fermi surface index.
In Eqs.~\eqref{Eq:eomr} and~\eqref{Eq:eomp}, ${\bf v}_{\varsigma{\bf p}}$ is the group velocity. 
Due to the ${\bf B}$ field, the orbital magnetic moment ${\bf m}_{\varsigma {\bf p} } = - i (e/2c) \left< \nabla_{\bf p} u_{\varsigma {\bf p}} \right| \times [\mathcal{H} - \mathcal{E}_{\varsigma {\bf p}  } ] \left| \nabla_{\bf p} u_{\varsigma {\bf p}} \right>$ modifies the energy as $\mathcal{E}_{\varsigma {\bf p} } \to \widetilde{\mathcal{E}}_{\varsigma {\bf p} }=\mathcal{E}_{\varsigma {\bf p} } - {\bf m}_{\varsigma {\bf p} } \cdot {\bf B}$ and ${\bf v}_{\varsigma{\bf p}}$ should be calculated from the modified energy $\widetilde{\mathcal{E}}_{\varsigma {\bf p} }$ as ${\bf v}_{\varsigma{\bf p}} = \partial \widetilde{\mathcal{E}}_{\varsigma} ({\bf p}) / \partial {\bf p}$.

We next attempt quantum analysis, motivated by the observation that the semiclassical analysis cannot capture quantum fluctuations.
For quantum analysis, we replace $\mathcal{H}$ in Eq.~\eqref{Eq:iso_Hamiltonian} with $\mathcal{H}_{L}$,
\begin{equation}
    \mathcal{H}_{L}  = \frac{1}{2 m^*}  
    \bm{\Pi}^2
    + \frac{\lambda}{\hbar}  
    \bm{\Pi} \cdot \bm{\sigma} 
    -\mu_{\rm B}^* {\bf B}\cdot \bm{\sigma},
    \label{Eq:LLHamil}
\end{equation}
where ${\bf \Pi}={\bf p}-e{\bf A}/c$ is the kinematic momentum operator in the presence of 
the vector potential ${\bf A}$ that is related to $\mathbf{E}=-(1/c)\partial \mathbf{A}/\partial t$ and $\mathbf{B}=\nabla \times \mathbf{A}$.  
The last term of $\mathcal{H}_{L}$ denotes the Zeeman coupling with the Bohr magneton $\mu_{\rm B}^*$. 
Then, we define the chirality operator $\hat{\mathcal{C}}$ as follows,
\begin{equation}\label{Eq:chirality}
	 \hat{\mathcal{C}}
	 =
	 -\text{sgn}\left[  \{ {\bf \Pi}_{\rm eff} \cdot {\bf v},  {\bf \Pi}_{\rm eff} \cdot \bm{\sigma} \}_{+}    \right],
\end{equation}
where ${\bf \Pi}_{\rm eff} = {\bf \Pi} + (\hbar / \lambda) ( e \hbar / 2 m^* c - \mu^*_{\rm B} ) {\bf B} $ is the effective kinetic momentum operator, ${\bf v}$ is the velocity operator, and $\{,\}_+$ denotes anticommutator. 
We note that Eq.~\eqref{Eq:chirality} differs from the helicity ${\bf \Pi} \cdot \bm{\sigma}$~\cite{PhysRevB.104.L241111,Huang2016}.
If the helicity is identified with the chirality~\cite{Huang2016}, ${\bf E} \cdot {\bf B}$ term in Eq.~\eqref{Eq:chiral_anomaly} vanishes when $\mathcal{E}_{\text{F}} < 0$.
Hereafter, we choose $\mu_{\rm B}^* =  e \hbar / 2 m^* c$ for simplicity, but this choice is not crucial. 
To demonstrate the physical meaning of $\hat{\mathcal{C}}$, we first consider the case $\mathbf{A}=0$ ($\mathbf{E}=\mathbf{B}=0$).
It is trivial to verify that $\hat{\mathcal{C}}$ has eigenvalues $\pm 1$ for the red/blue-colored states in Fig.~\ref{fig:Metal_B0}. 
Thus $\hat{\mathcal{C}}$ amounts to the chiral number operator (or quantum generalization of the Fermi surface index $\chi$ introduced in the above semiclassical analysis). 
This allows $n_\chi$ and $\mathbf{J}_\chi$ in Eq.~\eqref{Eq:chiral_anomaly} to be defined unambiguously in this quantum analysis.
Since $\hat{\mathcal{C}}$ and $\mathcal{H}_{L}$ share the complete set of common eigenstates, $[ \hat{\mathcal{C}}, \mathcal{H}_{L}] = 0$, which verifies the chiral number conservation for $\mathbf{E}=\mathbf{B}=0$. 
Interestingly, the commutator $[\hat{\mathcal{C}},\mathcal{H}_{\rm L}]$ vanishes even for nonzero $\mathbf{A}\neq 0$. 
If $\mathbf{A}$ is time-independent ($\mathbf{E}=0$), the vanishing commutator implies the chiral number conservation even for $\mathbf{B}\neq 0$.
If $\mathbf{A}$ is time-dependent ($\mathbf{E}\neq 0$), on the other hand, one obtains $d \hat{\mathcal{C}} / dt = [ \hat{\mathcal{C}}, \mathcal{H}_{\text{L}} ]/ i \hbar + \partial \hat{\mathcal{C}} / \partial t$, where $[ \hat{\mathcal{C}}, \mathcal{H}_{\text{L}} ] = 0$ and $\partial \hat{\mathcal{C}} / \partial t \neq 0$.
That is, the vanishing commutator does {\it not} necessarily imply the chiral number conservation.
%
%
The above operator analysis thus implies that nonvanishing $\mathbf{E}$ is a necessary condition for possible violation of the chiral number conservation.
%


However, nonvanishing $\mathbf{E}$ is not a sufficient condition.
To examine possible violation of the chiral number conservation, we calculate the $\mathbf{E}$-induced $\hat{\mathcal{C}}$ change, $\delta \langle \hat{\mathcal{C}} \rangle$ (${\bf B} = 0$), by using the Kubo formula, which takes into account both the inter- and intra-band contributions. 
The interband contribution $\delta\langle \hat{\mathcal{C}} \rangle_{\rm inter}$ is proportional to 
$\langle u_{\varsigma{\bf p}}| \hat{\mathcal{C}} |u_{\varsigma'{\bf p}}\rangle \langle u_{\varsigma'{\bf p}}|\mathbf{v}|u_{\varsigma{\bf p}}\rangle$ ($\varsigma' \neq \varsigma$), which vanishes since $\langle u_{\varsigma{\bf p}}| \hat{\mathcal{C}} |u_{\varsigma'{\bf p}}\rangle=0$ due to $[ \hat{\mathcal{C}} ,\mathcal{H}_{\rm L}]=0$.
Thus $\delta \langle \hat{\mathcal{C}} \rangle$ is determined by the interband contribution  $\delta \langle \hat{\mathcal{C}} \rangle_{\rm{intra}}$, which describes the Fermi surface shift. 
As shown in Fig.~\ref{fig:FSduetoE}, in the presence of the ${\bf E}$ field, there is no change of $\hat{\mathcal{C}}$ due to the Fermi surface shift by ${\bf E}$.
In other words, $\hat{\mathcal{C}}$ is preserved at each Fermi surface even though the ${\bf E}$ field is applied to the system.

If the ${\bf B}$ field is also present, $\delta \langle \hat{\mathcal{C}} \rangle_{\rm{intra}}$ can be nonzero as we demonstrate below.
When an external uniform ${\bf B}$ field along the $z$ direction is turned on, the Landau levels (LL) develop:
For $\nu \geq 1$, $\mathcal{E}_{\nu, \varsigma = \pm}(p_z) =   p_z^2/2m^* + \nu \hbar \omega_c \pm \Delta_{\nu}(p_z)$ and for $\nu$$=0$, $\mathcal{E}_{\nu=0}(p_z) =  p_z^2 / 2m^*  +   \lambda  p_z /\hbar $.
Here, $\Delta_{\nu}(p_z) = \sqrt{ 2 \nu \lambda^2 / l_B^2 +  \lambda^2 p_z^2 / \hbar^2 }$, $l_B = \sqrt{\hbar c/eB}$, and $\omega_c = e B / m^*c$.
The LL energy dispersion is shown in Fig.~\ref{fig:Metal_LLbandweakB}, where states with $\hat{\mathcal{C}}$ eigenvalue $+1 (-1)$ are colored in red (blue).
Note that the zeroth LL ($\nu = 0$) states have $\hat{\mathcal{C}} = -1$ for $p_z / \hbar \geq - m^* \lambda/ \hbar^2$ and $\hat{\mathcal{C}} = +1$ for $p_z / \hbar \leq  - m^* \lambda/ \hbar^2$, where $m^* \lambda/ \hbar^2 \simeq 0.13$ \AA$^{-1}$ [see vertical dashed line in Fig.~\ref{fig:Metal_LLbandweakB}].
Thus the state shift by the $z$ component of ${\bf E}$ can induce nonzero $\delta \langle \hat{\mathcal{C}} \rangle_{\rm{intra}}$.
For other LLs ($\nu \geq 1$), the state shift does {\it not} induce nonzero $\delta \langle \hat{\mathcal{C}} \rangle_{\rm{intra}}$ since the $\hat{\mathcal{C}}$ change at $p_z / \hbar \simeq  - m^* \lambda/ \hbar^2$ is counterbalanced by the $\hat{\mathcal{C}}$ change at $p_z / \hbar \simeq  m^* \lambda/ \hbar^2$.
Thus, when ${\bf E}$ is applied along the $z$ axis, we obtain nonzero $\delta \langle \hat{\mathcal{C}} \rangle  = \tau_{\rm{cv}}  e^2 {\bf E} \cdot {\bf B} /  2 \pi^2 \hbar^2 c $ due to the unbalance caused by the zeroth LL.
Here, $\tau_{\rm{cv}}$ is the chirality-violating scattering time [Fig.~\ref{fig:Metal_B0}].
$\delta \langle \hat{\mathcal{C}} \rangle$ agrees with the result of Eq.~\eqref{Eq:chiral_anomaly} for the chiral number density in the homogeneous system at steady state, $\partial n^{5} / \partial t  =  e^2 {\bf E} \cdot {\bf B} / 4 \pi^2 \hbar^2 c - \delta n^{5} / \tau_{\rm{cv}} = 0$ with $n^{5} = (n_{+1} - n_{-1})/2$.
This confirms the chiral anomaly for $\mathcal{H}_{L}$.

{\it Charge transport.---} 
It has been reported~\cite{PhysRevB.88.104412,Burkov:2015JPCM,PhysRevB.91.245157,PhysRevB.96.041110,PhysRevLett.119.176804} that the charge transport in Weyl semimetals acquires corrections to the Drude conductivity $\sigma_{\rm{D}}$ due to the chiral anomaly. 
Here we examine if the SOC-induced chiral anomaly affects the charge transport in a similar way. 
For this, we use the semiclassical expression for the charge current density $\bm{\mathcal{J}}$,  
\begin{equation}\label{Eq:totalJc}
     \bm{\mathcal{J}} =  e \sum_{\varsigma } \int \frac{ d^3 {\bf p}}{ ( 2 \pi \hbar)^3}~ \mathcal{G}_{\varsigma{\bf p}} \dot{\bf r}_{\varsigma} ~ f_{\varsigma {\bf p} },  
\end{equation}
which is valid when the Landau level quantization is not important.
Here $\mathcal{G}_{\varsigma{\bf p}} = 1 + e{\bf B} \cdot \bm{\Omega}_{\varsigma{\bf p}} /c $ is the Berry phase correction to the density of states~\cite{RevModPhys.82.1959} and the electron occupation function $f_{\varsigma{\bf p}}$ is given by
\begin{equation}\label{Eq:Boltzmann}
	f_{\varsigma {\bf p}}  
    = f_{\varsigma{\bf p}}^0 
	- \tau_{\text{cp}} ( \dot{\bf p} \cdot \nabla_{\bf p}) f_{\varsigma{\bf p} }^0  
	+ \mathcal{O}(\tau^2_{\rm{cp}}).	
\end{equation} 
Here $f_{\varsigma{\bf p}}^{0}$ is the equilibrium Fermi-Dirac distribution and $\tau_{\rm{cp}}$ is the momentum relaxation time for the chirality-preserving scattering [Fig.~\ref{fig:Metal_B0}]. 
Note that the difference between $f_{\varsigma{\bf p}}$ and $f_{\varsigma_{\bf p}}^0$ depends explicitly on $\tau_\text{cp}$ and $\tau_\text{cv}$ does not appear explicitly. 
The latter appears through the modification of $f^{0}_{\varsigma {\bf p}}$ via the shift of the chemical potential $\mu_{\chi}$ for the following reason.
The chiral anomaly tends to incresease $n_\chi$ by $k_\chi e^2 ({\bf E}\cdot {\bf B})/4\pi^2\hbar^2 c$ per second.
This tendency is counterbalanced by the chirality-violating scattering and at steady states, one obtains the density of the pumped electrons $\delta n^5= \delta n_{\chi=+1}=-\delta n_{\chi=-1}=\tau_{\rm cv}e^2 ({\bf E}\cdot{\bf B})/4\pi^2 \hbar^2 c$. 
This pumping results in the chemical potential imbalance between the two Fermi surfaces, $ \mu_{\chi=+1}-\mu_{\chi=-1} \equiv 2\mu^5$, where
\begin{equation}\label{Eq:muh}
	\mu^{5} =  \pi^2 \frac{ (\lambda^2 + \frac{ \hbar^2 \mathcal{E}_{\text{F}}}{ m^*}  ) \sqrt{ \lambda^2 + \frac{2 \hbar^2 \mathcal{E}_{\text{F}}}{ m^*}  } }{\mathcal{E}_\text{F}^2} \delta n^5 . 
\end{equation}
Here $\mathcal{E}_\text{F}$ is the Fermi energy in the absence of ${\bf E}$ and ${\bf B}$. Equation~\eqref{Eq:muh} is essentially identical to $\mu^5$ for a conventional Weyl system~\cite{PhysRevD.78.074033,li2016chiral}.

This $\mu^5$ affects the charge current in many ways since all $f_{\varsigma{\bf p}}^0$'s in the series expansion of $f_{\varsigma{\bf p}}$ [Eq.~\eqref{Eq:Boltzmann}] are affected by $\mu^5$~\cite{Shin:2017kk}.
First of all, the very first term $f_{\varsigma{\bf p}}^0$ in the series expansion generates an ``equilibrium" current density, 
\begin{equation}\label{Eq:chiral_magnetic}
    \bm{\mathcal{J}}^\text{CM}
    =
    \frac{e^2}{2 \pi^2 \hbar^2 c} \mu^5 {\bf B}
    =
    \frac{ 3 \mu_{\text{B}}^{*2} }{ 4 \mathcal{E}_{\text{F}}^{2}} \frac{ (4 \pi^2 \hbar^2 c / e^2 ) }{ \tau_{\text{cp}} } \delta n^{5}\sigma_{\text{D}} {\bf B},
\end{equation}
which is nothing but the chiral magnetic effect~\cite{Huang2016,PhysRevLett.116.077201} illustrated initially for the quark-gluon plasma with the chiral anomaly~\cite{PhysRevD.78.074033}.
Since $\mu^5$ ($\delta n^5$) is proportional to ${\bf E}\cdot{\bf B}$, $\mathcal{J}^\text{CM}$ produces the negative magnetoresistance when ${\bf E}$ and ${\bf B}$ are parallel~\cite{PhysRevB.88.104412,Burkov:2015JPCM,PhysRevB.91.245157}.
The second term $-\tau_\text{cp}(\dot{\bf p}\cdot \nabla_{\bf p})f_{\varsigma{\bf p}}^0$ in Eq.~\eqref{Eq:Boltzmann} also affects the charge transport and generates a nonlinear correction $\bm{\mathcal{J}}^\text{D-N}$~\cite{supp}, 
\begin{equation} \label{Eq:finalANL} 
\bm{ \mathcal{J} }^{\text{D-N}}   
	= -    \frac{ \hbar^4 /m^{*2} }{ \lambda^2 + \hbar^2 \mathcal{E}_{\text{F}} / m^*} 	\frac{ \lambda }{2 \mathcal{E}_{  \text{F}}}  \delta n^5 
\sigma_\text{D}{\bf E},
\end{equation}
which is quadratic in ${\bf E}$ (note that $\delta n^5$ is also linear in ${\bf E}$) and makes the electric response nonreciprocal~\cite{Tokura:2018jd}.
A similar quadratic-in-${\bf E}$ correction is reported~\cite{PhysRevLett.117.146603} in noncentrosymmetric Weyl semimetals with {\it four} Weyl points where it is crucial for two Weyl points to be shifted in energy from the other Weyl points for such a nonlinear correction to arise.
In contrast, $\mathcal{H}$ has only one Weyl point.
We briefly mention that the energy correction due to ${\bf B}$ is ignored for $\mathcal{J}^{\rm{CM}}$ and $\mathcal{J}^{\rm{D-N}}$ since its effect is of higher order, $\mathcal{O}(B^3)$.

\begin{figure}[t!]     
  \includegraphics[width=8cm,height=2.96cm]{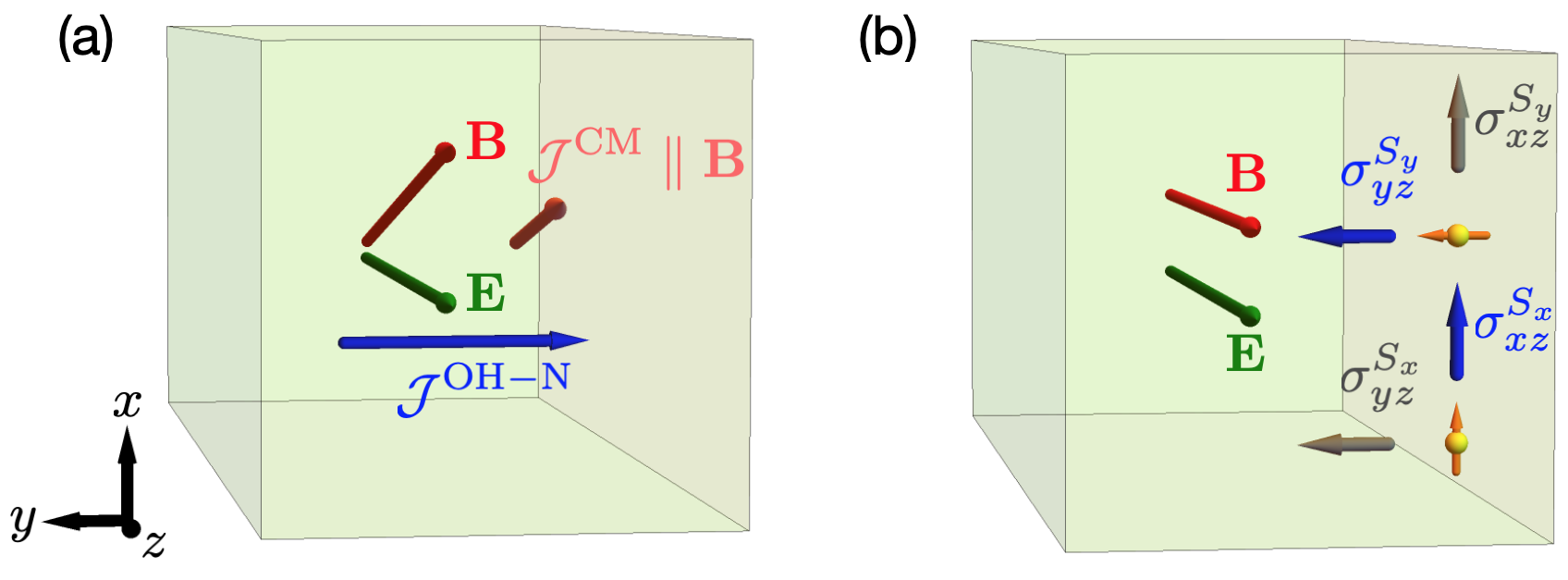}
   \subfigure{\label{fig:OH-N}}
   \subfigure{\label{fig:SHE}} 
\caption 
   {
   Schematic illustrations of (a) charge currents and (b) spin currents in the presence of both ${\bf E}$ and ${\bf B}$ fields.
   }
\label{fig02}
\end{figure}

The charge pumping affects not only the longitudinal current but also the Hall current, generating corrections to the Lorentz-force-induced Hall current, which we do not specify, and the orbital magnetic moment-induced Hall current~\cite{PhysRevB.103.125432} $\bm{\mathcal{J}}^{\text{OH}} = (e^3 / 24 \pi^2 \hbar^2 c ) \mathcal{L}_{\text{F}} {\bf B} \times {\bf E}$, where $\mathcal{L}_{\text{F}} = \lambda^2  /  ( \mathcal{E}_{\text{F}} \sqrt{\lambda^2 + 2 \hbar^2 \mathcal{E}_{\text{F}} / m^* })$.
Actually, the ``equilibrium" current density $\bm{\mathcal{J}}^\text{CM}$ already contains such a correction. 
Since $\bm{\mathcal{J}}^{\rm{CM}} \propto ({\bf E} \cdot {\bf B}) {\bf B}$ generates the planar Hall effect when ${\bf B}$ is not parallel to ${\bf E}$ but not perpendicular either [Fig.\ref{fig:OH-N}] as demonstrated previously for Weyl semimetals~\cite{PhysRevB.96.041110,PhysRevLett.119.176804}.  
There is another Hall current contribution arising from $f^{0}_{\varsigma{\bf p}}$ since the energy $\widetilde{\mathcal{E}}_{\varsigma{\bf p}}$ in the presence of ${\bf B}$ contains the correction $ - {\bf m}_{\varsigma {\bf p}} \cdot {\bf B}$.
Then, the resulting occupation correction $f^{0}_{\varsigma {\bf p} } ( \widetilde{\mathcal{E}}_{\varsigma {\bf p} } ) - f^{0} (\mathcal{E}_{ \varsigma {\bf p} } ) = - {\bf m}_{\varsigma {\bf p}  } \cdot {\bf B}  \partial f^{0}_{\varsigma {\bf p} } / \partial \mathcal{E}_{ \varsigma {\bf p} }    + \mathcal{O}(B^2)$ generates the Hall current correction $\mathcal{J}^\text{OH-N}$~\cite{supp}, 
\begin{equation}\label{Eq:OH-N}
    \bm{\mathcal{J}}^\text{OH-N}= \xi \frac{e \mu_\text{B}^{*2} \lambda }{16 \mathcal{E}_\text{F}^4}  \frac{ (4 \pi^2 \hbar^2 c/e^2 )  }{\tau_\text{cp}} \delta n^{5}\sigma_\text{D}  {\bf B}\times{\bf E},
\end{equation}
where the coefficient $\xi = 1 + \lambda^2 ( \lambda^2 + 3 \hbar^2 \mathcal{E}_{\text{F}}/ m^* ) / [( \lambda^2 +   \hbar^2 \mathcal{E}_{\text{F}} / m^*)(  \lambda^2 + 2 \hbar^2 \mathcal{E}_{\text{F}}/ m^*) ]$ approaches 1 as $\mathcal{E}_\text{F}\rightarrow 0$. 
In Fig.~\ref{fig:OH-N}, the blue arrow represents $\mathcal{J}^\text{OH-N}$.
The factor $\delta n^{5}$ in $\bm{\mathcal{J}}^\text{OH-N}$ indicates that $\bm{\mathcal{J}}^\text{OH-N}$ arises from the chiral anomaly and thus differs from the nonlinear Hall effect demonstrated for a tilted Weyl semimetal~\cite{li2020chiral} and a multi-Weyl semimetal~\cite{PhysRevB.104.205124}, which are not due to the anomaly-induced pumping $\delta n^{5}$.
%

{\it Spin transport.---}
Electron spin is an essential ingredient of the SOC-induced chiral anomaly. 
It is thus natural to expect that the chiral anomaly affects spin transport as well. 
In contrast, the spin is not essential in conventional realization of the chiral anomaly in Dirac and Weyl semimetals, and thus the chiral anomaly effect on the spin transport has not been examined except for Ref.~\cite{gao2021chiral} that examines the spin current injection effect.
Here we examine the chiral anomaly effect on the spin transport in the linear response regime.

First of all, we introduce the spin current operator $\hat{\mathcal{J}}^{S_{j} }_{i} = \frac{1}{2} \{  \partial \mathcal{H}/\partial p_{i}  , \sigma_{j}  \}$~\cite{RevModPhys.87.1213}, where the curly bracket denotes an anticommutator.
In the linear response regime, its expectation value $\mathcal{J}_{i}^{S_{j} }$ is proportional to ${\bf E}$, $\mathcal{J}_{i}^{ S_{j} }=\sigma_{i k}^{ S_{j} } E_k$, where the spin conductivity $\sigma_{i k}^{S_{j} }$ may depend on ${\bf B}$. 
Since we are interested in the chiral anomaly effect, we assume for simplicity that ${\bf E}$ and ${\bf B}$ are parallel to each other, both applied along the $z$-axis. 
Then the relevant spin conductivity matrix $\sigma_{ i z}^{ S_j }$ has the following structure,
\begin{equation}\label{Eq:JsCom}
	 \sigma^{S_{j} }_{i z} 
	 =
	 \begin{pmatrix}
	 	\sigma_{xz}^{S_x}  & \sigma_{xz}^{S_y}  & 0 \\
	 	\sigma_{yz}^{S_x}  & \sigma_{yz}^{S_y} & 0 \\
	 	0 & 0 & \sigma_{zz}^{S_z}
	 \end{pmatrix}, 
\end{equation} 
where $\sigma_{xz}^{S_z}=\sigma_{yz}^{S_z}=\sigma_{zz}^{S_x}=\sigma_{zz}^{S_y}=0$ due to the rotation symmetry of $\mathcal{H}$. 
The same symmetry also guarantees $\sigma_{xz}^{S_x}=\sigma_{yz}^{S_y}$ and $\sigma_{xz}^{S_y} = -\sigma_{yz}^{S_x}$~\cite{PhysRevResearch.2.023065}.
In Fig.~\ref{fig:SHE}, gray and blue arrows represent the spin current.
In Eq.~\eqref{Eq:JsCom}, the off-diagonal components $\sigma_{xz}^{S_y} = -\sigma_{yz}^{S_x}$ are the conventional spin Hall conductivity~\cite{RevModPhys.87.1213}. 
The spin current described by the diagonal components $\sigma_{xz}^{S_x} = \sigma_{yz}^{S_y}$ is similar to the magnetic spin Hall current~\cite{Kimata:2019cq} in ferromagnetic systems~\cite{PhysRevResearch.2.023065} except that ${\bf B}$ in our system plays the role of the ferromagnetic magnetization ${\bf M}$ in magnetic spin Hall systems. 

The chiral anomaly effect on the spin transport in the linear response regime can be captured by considering the pumping $\delta n^5$.  
From $\mathcal{J}^{S_{j}}_{i} = \sum_{\varsigma} (2 \pi \hbar)^{-3}  \int d^3{\bf p} ~  \mathcal{G}_{\varsigma{\bf p}} \langle u_{\varsigma{\bf p}} | \hat{\mathcal{J}}^{S_{j} }_{i} | u_{\varsigma{\bf p}} \rangle [f_{\varsigma{\bf p}}^0(\delta n^5) - f_{\varsigma{\bf p}}^0(\delta n^5=0)]$, we obtain the anomaly-induced spin current $\mathcal{J}_x^{S_x}=\mathcal{J}_y^{S_y}=\mathcal{J}_z^{S_z}=\mathcal{J}_s^\text{ano}$, where $\mathcal{J}_s^\text{ano}$ is given by
\begin{equation}\label{Eq:AnoJSpin}
	\mathcal{J}^{\text{ano}}_{s}
	= 
	-  \frac{ \mu_{\text{B}}^{*} }{ 2 ( \mathcal{E}_{\text{F}} + m^* \lambda^2 / \hbar^2 )} 
	     \frac{ (4 \pi^2 \hbar^2 c / e^2 ) }{ \tau_{\text{cp}} } 
	     \delta n^5
	     \frac{ \sigma_{\text{D}} }{ e }.
\end{equation}  
Note that $\mathcal{J}^{\rm{ano}}_{s}$ contributes only the diagonal spin conductivities $\sigma_{xz}^{S_x} = \sigma_{yz}^{S_y} = \sigma_{zz}^{S_z}$.

We estimate the magnitudes of various anomaly-induced charge/spin currents, all of which increase as $\mathcal{E}_{\rm{F}}$ approaches the Weyl point energy, $0$.
This increase is partially due to the explicit inverse proportionality to $\mathcal{E}_{\rm{F}}$ [Eqs.~\eqref{Eq:chiral_magnetic},~\eqref{Eq:finalANL},~\eqref{Eq:OH-N},~\eqref{Eq:AnoJSpin}] and also due to the implicit inverse proportionality to $\mathcal{E}_{\rm{F}}$ hidden in $\delta n^{5} \propto \tau_{\rm{cv}}$.
After some calculation~\cite{supp}, one obtains $\tau_{\rm{cv}}/\tau_{\rm{cp}} \sim (2 m^* \lambda^2 / \hbar^2)^2 / \mathcal{E}_{\rm{F}}^2$~\cite{scatteringrate}.
For the parameters, $m^* = 1.4 m_e$, $\lambda = 0.7$ eV$\cdot$\AA, $\mathcal{E}_{\text{F}} = 2$ meV, $\tau_{\rm{cp}} = 6.85$ ps, and for the strengths $E= 10^{4} \text{ A/m}$ and $B = 1$ T, we obtain $\mathcal{J}^{\rm{D}} \simeq 2 \times 10^7 \rm{A/cm}^2$.
And, $\mathcal{J}^{\text{CM}} / \mathcal{J}^{\text{D}} \simeq 260$ \%, $\mathcal{J}^{\text{D-N}} / \mathcal{J}^{\text{D}} \simeq 32$ \%, $\mathcal{J}^{\text{OH-N}} / \mathcal{J}^{\text{D}} \simeq 7$ \%, and $\mathcal{J}^{\text{ano}}_s / \mathcal{J}^{\text{D}} \simeq 180$ \%. 
We note however that these ratios decrease significantly when $|\mathcal{E}_{\rm{F}}|$ becomes larger.

Lastly we remark that the SOC-induced chiral anomaly is not limited to the Weyl-type SOC ${\bf p} \cdot \bm{\sigma}$ in the point groups ${\bf T}$ and ${\bf O}$.
The SOC structure varies with the crystal structure~\cite{KVSamokhin:2009dg} and we verify~\cite{supp} that the SOC structure in 11 out of the total of 21 noncentrosymmetric point groups can generate nonvanishing Berry curvature flux through Fermi surfaces.
Although details of the anomaly-induced charge and spin transport properties may be affected by the details of the SOC structure, our work demonstrates the abundance of the SOC-induced chiral anomaly.

To conclude, we demonstrated that the SOC-induced chiral anomaly can occur in diverse noncentrosymmetric systems even without pairs of Weyl points and without the chiral symmetry, provided the Berry curvature flux through Fermi surfaces is nonzero.
This motivates the reinterpretation of the condensed matter chiral anomaly as a {\it Fermi surface} property rather than a Weyl point property, as proposed initially by Ref.~\cite{PhysRevLett.109.181602}.


\begin{acknowledgments}
S. Cheon thanks I. Jang for useful discussion.
H. -W. Lee and S. Cheon were supported by the National Research Foundation (NRF) of Korea (Grant No. 2020R1A2C2013484).
G.Y.C.  acknowledges the support of the NRF of Korea funded by the Korean Government (Grants No. 2020R1C1C1006048 and No. 2020R1A4A3079707).
G.Y.C is also partially supported by IBS-R014-D1.
G.Y.C. is also supported by the Air Force Office of Scientific Research under Award No. FA2386-20-1-4029.
G.Y.C also acknowledges financial support from Samsung Science and Technology Foundation under Project Number SSTF-BA2002-05.
K.-S. Kim was supported by the Ministry of Education, Science, and Technology (NRF-2021R1A2C1006453 and NRF-2021R1A4A3029839) of the National Research Foundation of Korea (NRF).
\end{acknowledgments}


\begin{thebibliography}{49}
\bibitem [{sca()}]{scatteringrate}%
  \BibitemOpen
  \href@noop {} {}\bibinfo {note} {When $\mathcal{E}_{\rm F}$ is large (away
  from the Weyl point) or $\lambda$ is small (weak SOC), however,
  $\tau_{\text{CP}}\simeq \tau_{\text{CV}}$.}\BibitemShut {Stop}%
\end{thebibliography}

%

\end{document}